\documentclass[twocolumn,prb]{revtex4}

\usepackage{graphicx}
\usepackage{dcolumn}
\usepackage{bm}
\usepackage[cmex10]{amsmath}

\begin{document}

\preprint{APS/123-QED}

\title{Spin transfer torque oscillator based on asymmetric magnetic tunnel junctions}

\author{Witold Skowro\'{n}ski}
 \email{skowron@agh.edu.pl}
\author{Tomasz Stobiecki}
\author{Jerzy Wrona}
\affiliation{Department of Electronics, AGH University of Science and Technology, Al. Mickiewicza 30, 30-059 Krak\'{o}w, Poland
}

\author{G\"{u}nter Reiss}
\affiliation{Thin Films and Physics of Nanostructures, Bielefeld University, 33615 Bielefeld, Germany}

\author{Sebastiaan van Dijken}
\affiliation{NanoSpin, Department of Applied Physics, Aalto University School of Science, P.O.Box 15100, FI-00076 Aalto, Finland}

\date{\today}

\begin {abstract}
We present a study of the spin transfer torque oscillator based on CoFeB/MgO/CoFeB asymmetric magnetic tunnel junctions. We observe microwave precession in junctions with different thickness of the free magnetization layer. Taking advantage of the ferromagnetic interlayer exchange coupling between the free and reference layer in the MTJ and perpendicular interface anisotropy in thin CoFeB electrode we demonstrate the nanometer scale device that can generate high frequency signal without external magnetic field applied. The amplitude of the oscillation exceeds 10 nV/$\sqrt{Hz}$ at 1.5 GHz. 
\end{abstract}

\pacs{75.47.-m, 72.25.-b}
\maketitle

Magnetic tunnel junctions (MTJs) consisting of two ferromagnetic electrodes separated by a thin tunnel barrier has recently drawn a significant attention due to their potential applications as a high density memory cell \cite{huai_observation_2004, takemura_32-mb_2010} and microwave electronic components \cite{kiselev_microwave_2003, rippard_direct-current_2004, deac_bias-driven_2008}. DC currents in such structures can induce steady state precessions due to the interaction between spin-polarized electrons and the local magnetization of the free layer (FL). This spin-transfer-torque (STT) effect \cite{slonczewski_current-driven_1996, berger_emission_1996} induces resistance fluctuations in the MTJ which in turn generate an AC signal in the GHz frequency range. Such STT-based nanometer scale oscillator can be a competitive device to the existing LC-tank technologies used widely in high-frequency electronics. One of the key issues of the spin torque oscillators (STOs) is the ability to produce microwave signal without the a need of operating in an external magnetic field. In this work, we report on STO based on asymmetric magnetic tunnel junctions with the thin MgO tunnel barrier and the FL ferromagnetically coupled to the reference layer (RL) that are able to operate with no magnetic field applied. To our knowledge, such operation has not been published yet.

The MTJ stack with a CoFeB wedged shaped electrode was deposited in a Singulus Timaris cluster tool system. The multilayer structure consisted of the following materials (thickness in nm): buffer layers / PtMn (16) / Co$_{70}$Fe$_{30}$(2) / Ru(0.9) / Co$_{40}$Fe$_{40}$B$_{20}$(2.3) /  MgO(0.85) / Co$_{40}$Fe$_{40}$B$_{20}$(1 - 2.3) / capping layer. The deposition process was similar to the one used in our previous studies \cite{skowronski_interlayer_2010, wrona_low_2010}. After deposition, three different parts of the sample with different FL thickness were selected for patterning into nanometer size pillars (later in the paper referred to as A1, A2 and A3, see Table ~\ref{tab:Static} for details). In this paper we focus mainly on the sample with 1.57 nm thick FL - A2. Using a three-steps electron beam lithography process, which included ion beam milling, lift-off and oxide deposition steps, nanopillars with elliptical cross-section of 250 $\times$ 150 nm were fabricated. The pillars were etched down to the PtMn layer. To ensure good RF performance of the device, the overlap between the top and bottom leads was about 4 $\mu$m$^2$, which resulted in a capacitance of less than 1 $\times$ 10$^{-14}$ F.
The DC measurements were conducted at room temperature with a magnetic field applied in the sample plane. The high-frequency measurements were carried out using a Agilent N9030A spectrum analyzer with built preamplifier. The MTJ bonded to the high frequency chip carrier was connected to a bias-tee. The DC signal from a sourcemeter was fed to the sample through the inductive connector of the bias-tee, whereas the spectrum was measured at the capacitive connector. In this paper, the positive voltage denotes the electron flowing from a bottom RL to the top FL favoring the parallel alignment of the magnetizations.

\begin{figure}
\includegraphics[width=2.5in]{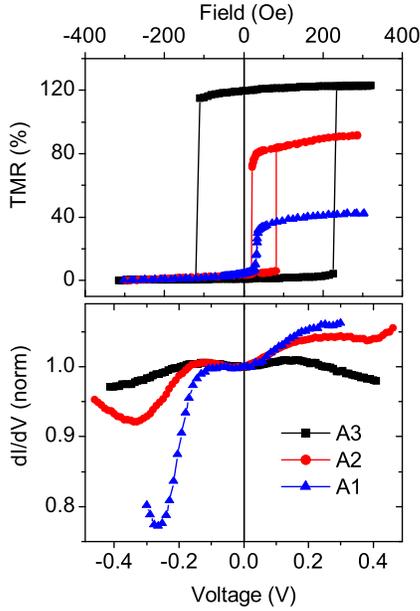}
\caption{TMR vs.  magnetic field (a) and differential conductance (dI/dV) vs. DC bias voltage at low-resistance state (b) measured for samples with different FL thickness. Clear asymmetry between CoFeB electrodes is observed in dI/dV measurement for samples A1 and A2.}
\label{fig:TMR_cond}
\end{figure}


Figure \ref{fig:TMR_cond}a shows the TMR loops for samples A1-A3, with field applied along the in-plane easy axis of the MTJ. The TMR decreases with decreasing FL thickness due to reduced spin polarization of the tunneling electrons \cite{sun_dependence_1999}. Moreover, the magnetization of the FL is tilted out of the film plane due to the perpendicular interface anisotropy in thin CoFeB layer \cite{ikeda_perpendicular-anisotropy_2010, khalili_amiri_switching_2011} and therefore, at zero magnetic field a full parallel state for samples A1 and A2 is not achieved. The coercive field of about 100 Oe for sample A3 is reduced to zero for sample A1 \cite{wisniowski_effect_2008}. Differential conductance versus DC bias voltage was measured for all samples using lock-in technique. The results are presented in Fig. \ref{fig:TMR_cond}b. The asymmetry between the thin FL and the RL for samples A2 and A3 is observed in comparison with symmetric A3 sample. This asymmetry arises from a different band structures in the ferromagnetic electrodes \cite{oh_bias-voltage_2009}. 

\begin{table}[!t]
\renewcommand{\arraystretch}{1.3}
\caption{Summary of static parameters of the prepared MTJ nanopillars.}
\label{tab:Static}
\centering
\begin{tabular}{cccc}
Sample No. & FL thickness & TMR & Ic P $\rightarrow$ AP  \\
  & (nm) & (\%) & mA  \\
\hline
A1 & 1.35 & 50 &  -0.95\\
A2 & 1.57 & 100 & -1.8 \\
A3 & 2.3 & 120 & -2.4\\
\end{tabular}
\end{table} 

\begin{figure}
\includegraphics[width=2.5in]{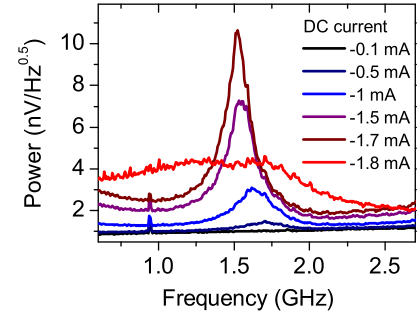}
\caption{Sample STO spectra of A2 measured at low-resistance state and different negative current applied to the MTJ without external magnetic field. Peak oscillation amplitude exceeds 10 nV/$\sqrt{Hz}$ at -1.7 mA. Further increase of current magnitude results in switching the MTJ to the high-resistance state.}
\label{fig:Spectrum_A2}
\end{figure}

Ferromagnetic coupling between the RL and FL stabilizes the low resistance state of the MTJs at zero applied magnetic field \cite{skowronski_interlayer_2010, serrano-guisan_inductive_2011}. All samples exhibited clear current induced magnetization switching measured for relatively long current pulses of 10 ms. The absolute switching current value needed to change the MTJ state from a P to AP, measured with no magnetic field assistance, was found to decrease with FL thickness due to reduced saturation magnetization (volume) of the FL (Table \ref{tab:Static}). 
The sample spectra of A2 measured with no magnetic field applied are shown in Fig. \ref{fig:Spectrum_A2}. Existence of the perpendicular interface anisotropy in the FL results in non zero angle theta between FL and RL magnetizations in a low resistance state at zero magnetic field (Fig. \ref{fig:TMR_cond}) and therefore the STT precession is excited even at low DC bias. An increase of the negative DC current of the polarization that favor the AP state (electrons flowing from the FL to the RL) results in increased amplitude of the oscillations, that exceeds 10 nV/$\sqrt{Hz}$ at 1.5 GHz and -1.7 mA. Further increase of the negative current magnitude results in switching the MTJ to the high resistance AP state, where peak of much smaller amplitude and wider linewidth is observed.

\begin{figure}
\includegraphics[width=2.5in]{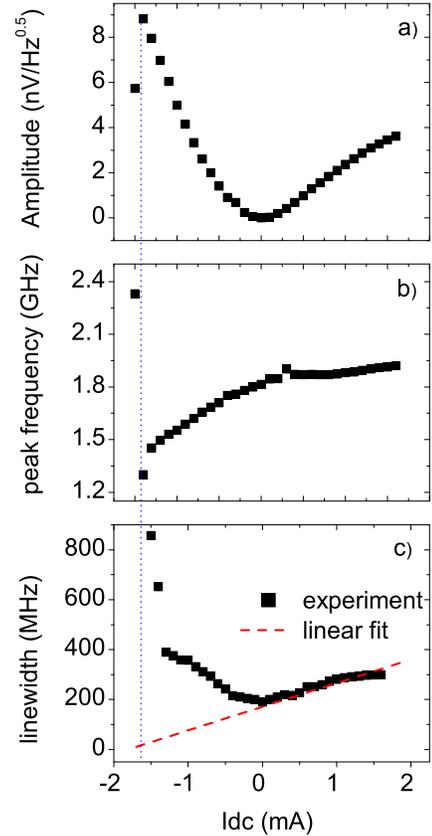}
\caption{The DC bias current dependence of the amplitude (a), peak frequency \textit{f$_{0}$} (b) and linewidth $\Delta$\textit{f} (c). No magnetic field was applied during the measurements. The dashed line in (c) represents the linear fit to the experimental data for positive I$_{dc}$. Near the switching current (dotted line) both \textit{f$_{0}$} and $\Delta$\textit{f} increase.}
\label{fig:A2_sum}
\end{figure}

Fig. \ref{fig:A2_sum} presents the STO amplitude, peak frequency (\textit{f$_{0}$}) and linewidth $\Delta$\textit{f} versus DC bias current with no magnetic field applied. Clearly, the oscillations for negative current favoring the AP state are more powerful than for the positive one, favoring P state, however peak at both current polarizations are visible due to a non zero angle theta at zero field. The dependence of linewidth of DC current is expressed as: 
\begin{equation}
\label{eq:linewidth}
\Delta f = \frac{\sigma}{2\pi}\left(I_{c0}-I_{dc}\right)
\end{equation} 
where $\sigma$ is the spin polarization efficiency and \textit{I$_{c0}$} is the threshold current \cite{wada_spin-transfer-torque-induced_2010}. Extrapolating the $\Delta$\textit{f} at the damping side (when MTJ is in P state and current favors P state) to zero Hz estimates the threshold current value (dotted line in Fig. \ref{fig:A2_sum}. Moreover, a rapid change both in \textit{f$_{0}$} and $\Delta$\textit{f} is observed near the switching threshold. A similar current value was measured during the static CIMS experiment. The switching voltage is much smaller than the breakdown voltage, therefore we can induce steady state precession without destroying the MTJ.

\begin{figure}[!t]
\includegraphics[width=2.5in]{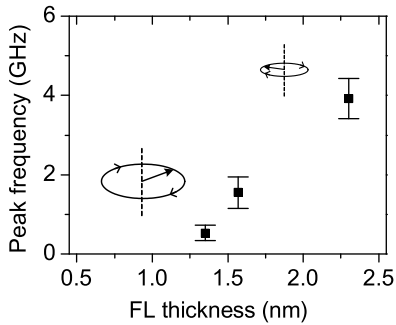}
\caption{Peak frequencies for MTJs with different thickness of the FL measured at low resistance state and \textit{I$_{dc}$} = -1 mA without magnetic field applied.}
\label{fig:Peak_freq_sum}
\end{figure}

The peak frequency at constant \textit{I$_{dc}$} = -1 mA and (\textit{H$_{dc}$}) = 0 was found to increase with increasing FL thickness, results are presented in Fig. \ref{fig:Peak_freq_sum}. Increased anisotropy constant results in smaller precession trajectories and therefore increased \textit{f$_{0}$}. It should be noted, that a sample-to-sample distribution in both oscillation's amplitude and frequency is observed, mainly due to the size and shape distribution during a nano-lithography process, however the overall tendency is retained. For other samples with thinner FL of 1.22 nm we were not able to observe any oscillation in the measured bandwidth even with strong magnetic field applied in-plane of perpendicular to the sample's easy axis. 
For sample A2 the oscillation's amplitude is of the same order than the highest reported to date \cite{deac_bias-driven_2008} exceeding 10 nV/$\sqrt{Hz}$. We conclude, that taking advantage of the coupling mechanisms in MTJs with a thin MgO tunnel barrier in combination with perpendicular interface anisotropy might enhance STOs performance without the need of an external magnetic field application. 

In summary we have demonstrated an STO based on an asymmetric MTJ, that is able to produce a microwave signal without a need of the magnetic field applications. Due to the ferromagnetic interlayer exchange coupling in our system, the MTJ is in stable low resistance state at \textit{H$_{dc}$} = 0. Perpendicular interface anisotropy is thin FL is used, to induce the magnetization precession at small DC bias. The oscillation's amplitude exceeds 10 nV/$\sqrt{Hz}$ at 1.5 GHz and \textit{I$_{dc}$} = -1.7 mA.

The authors would like to thank Singulus Technologies AG for consultation and technical help with MgO wedge MTJs preparation. Work supported by the SPINLAB  POIG.02.02.00-00-020/09 project. T.S. and W.S. acknowledge Foundation for Polish Science MPD Programme co-financed by the EU European Regional Development Fund. and the Polish Ministry of Science and Higher Education grants (IP 2010037970 and NN 515544538). S.v.D. acknowledges financial support from the Academy of Finland for the ACTIVE-BAR project (no. 127731).

\bibliographystyle{abbrv}
\bibliographystyle{unsrt}
\bibliography{Skowronski_library}

\end{document}